\documentstyle[11pt,epsfig]{article}
\author{M. Mohseni\thanks{email: mohseni@cc.sbu.ac.ir}\hspace{1mm}
and \hspace{1mm}H. R. Sepangi\thanks{ email:
hr-sepangi@cc.sbu.ac.ir}
\\ \small {Physics Department, Shahid
Beheshti University, Evin, Tehran 19839, Iran}}
\begin{document}
\title{\bf Gravitational waves and spinning test particles}
\maketitle
\begin{abstract}
The motion of a classical  spinning test particle in the field of
a weak plane gravitational wave is studied. It is found that the
characteristic dimensions of the particle's orbit is sensitive to
the ratio of the spin to the mass of the particle. The results are
compared with the corresponding motion of a particle without
spin.\vspace{5mm}
\end{abstract}
\section{Introduction}
One of the problems of great importance in contemporary physics is
the detection of gravitational waves which would profoundly affect
both physics and observational astronomy \cite{sc}. Like other
kind of  waves, one may think of gravitational waves as wiggles
propagating in some background. In fact these waves are wiggles of
curvature propagating in a fixed background with the speed of
light. The corresponding space-time metric may be written as
$g_{\mu\nu}=g^0_{\mu\nu}+h_{\mu\nu}$ in which $g^0_{\mu\nu}$
stands for the background metric and $h_{\mu\nu}$ represents the
gravitational wave field. In practical situations where the
detection of these waves is our prime concern, one needs to
consider only weak, plane, linear, waves in an essentially flat
background because the amplitude of the waves crossing the solar
system are very small even though the energy flux may be enormous
\cite{b}. We are thus led to the  decomposition of the space-time
metric as $g_{\mu\nu}=\eta_{\mu\nu}+h_{\mu\nu}$ where
$\eta_{\mu\nu}=(-,+,+,+)$ is the usual Minkowski metric and
$|h_{\mu\nu}|\ll{1}$. It then follows from general relativity that
$h_{\mu\nu}$ satisfies a generalized wave equation in three
dimensions and there are only two components for $h_{\mu\nu}$
corresponding to two independent polarizations of gravitational
waves usually denoted by $h_{+}$ and $h_{\times}$ \cite{m}.
Polarization of these waves contains valuable information about
the sources generating them \cite{sc}.

Gravitational waves alter the curvature of the region of the
space-time they cross and hence affect the motion of particles.
This problem has been studied extensively in the past {\it e.g.}
see \cite{b}-\cite{v}. The basic idea is that free particles move
on the geodesics of the basic field $g_{\mu\nu}$. The physical
effect of a passing gravitational wave is that particles acquire a
relative acceleration during the passage of the wave and hence
gain a relative velocity after the wave has passed. The
measurement of this relative velocity would hopefully provide a
way to detect gravitational waves.

Grishchuck \cite{g} studied the problem of the motion of a
particle in the field of weak plane gravitational waves with
different polarizations and calculated the trajectory of the
particle as viewed from a locally inertial frame, that is, from
the reference frame attached to a nearby particle. These
trajectories, when projected onto the plane perpendicular to the
direction of the wave propagation, become ellipsis or segments of
straight lines whose dimension and orientation depend on the
initial position of the particle and the polarization of the wave.
By a suitable combination of the particle's initial position and
wavelength of the wave, it is possible to find invariant
dimensionless parameters which characterize the dimension of the
trajectories.

In this paper we consider the problem of the determination of the
motion of a classical free spinning particle in the field of a
weak, plane  gravitational wave with arbitrary polarizations.
Here, several interesting phenomena will appear. The particle's
spin couples to gravity and this makes the spin dynamical. Also,
the trajectory of the particle deviates from geodesics of the
field $g_{\mu\nu}$ as a result of a Lorentz-like force which
originates from the spin-gravity coupling. This problem was
previously considered from a different point of view in \cite{bi}
where a supplementary equation was proposed causing the spinning
particle to move along a geodesic.

In what follows we first review the equations of motion of a
spinning particle in a general space-time, sometimes known as Dixon's equations
\cite{eh}, and derive their  weak field
approximation. We then apply them to the case of a spinning test particle in
the field of a gravitational wave and compare the results with those of a
test particle without spin.
\section{Dixon's equations}
The motion of an extended body in a given background may be
studied by treating the body via a multipole expansion. This
expansion is based on the following procedure. The body sweeps out
a narrow tube in space-time as it moves. Let $L$ be a line inside
this tube representing the motion of the body. Denote the
coordinates of the points on this line by $X^{\alpha}$ and define
${\delta}x^{\alpha}=X^\alpha-x^\alpha$, with $x^{\alpha}$ being
the coordinates of the points of the body. Let $T_{\mu\nu}$ be the
energy-momentum tensor describing the body and consider the
quantities $\int{T^{\mu\nu}dV}$, $\int{{\delta}x^
{\lambda}T^{\mu\nu}}dV$,
$\int{{\delta}x^{\lambda}{\delta}x^{\rho}T^{\mu\nu}}dV, \cdots$
where the integrations are carried out on the three-dimensional
hypersurfaces of fixed time $X^0=\mbox{const.}$. These are in fact
the successive terms in the above mentioned multipole expansion.
If we only keep the first term, that is, if we assume that all the
integrals containing ${\delta}x^\mu$ vanish, this will describe a
single-pole particle. Similarly, we may consider a pole-dipole
particle for which all the integrals with more than one factor of
${\delta}x^\mu$ vanish. Higher  order approximations may be
defined in a similar way. In the limit of $||{\delta}x^{\alpha}||
{\rightarrow}0$ with $||.||$ denoting the norm with respect to the
induced metric on the three-dimensional hypersurface of constant
time, the approximation becomes exact. Thus a single-pole particle
is in fact a test particle without any internal structure. A
pole-dipole particle is a test particle whose internal structure
is expressed by its spin, an antisymmetric second rank tensor
defined by
\begin{eqnarray*}
s^{\mu\nu}{\equiv}\int\left(\delta x^{\mu}T^{0\nu}-\delta
x^{\nu}T^{0\mu}\right)dV.
\end{eqnarray*}
The equations of motion are then obtained by applying
Einstein's field equations together with the conservation of the
energy-momentum tensor, $\nabla_{\mu}T^{\mu\nu}=0$, describing the body.
For a single-pole this leads to a free particle moving along the
geodesics associated with $g_{\mu\nu}$. This is the same as that
obtained by Einstein and co-workers, who only made use of the field equations
without resorting to the dynamical equation. Using the
above procedure, Papapetrou \cite{pa} derived two sets of equations
for the motion of a pole-dipole particle and its spin
\begin{equation}
{\dot{p}}^{\mu}=
-\frac{1}{2}R^{\mu}_{\hspace{2mm}\nu\lambda\rho}v^{\nu}
s^{\lambda\rho}\label{eq1}
\end{equation}
\begin{equation}
{\dot{s}}^{\mu\nu}= p^{\mu}v^{\nu} - p^{\nu}v^{\mu} \label{eq2}
\end{equation}
in which $v^{\mu}=\frac{dX^{\mu}(\tau)}{d\tau}$ is the
four-velocity of the particle along $L$ parameterized by $\tau$
with $v\cdot v=-1$, $ R^{\mu}_{\hspace{2mm}\nu\lambda\rho}$ is the
curvature tensor, $p^{\mu}=\int{T^ {0\mu}dV}$ represents the
particle's four-momentum along $L$ and
\begin{eqnarray*}
{\dot{p}}^{\mu}&{\equiv}&\frac{dp^{\mu}(\tau)}{d\tau}+
{\Gamma}^{\mu}_{\hspace{2mm}
\lambda\nu}v^{\lambda}p^{\nu}\\{\dot{s}}^{\mu\nu}&{\equiv}&
\frac{ds^{\mu\nu}(\tau)}
{d\tau}+{\Gamma}^{\mu}_{\hspace{2mm}\lambda\rho}v^{\lambda}s^{\rho\nu}+
{\Gamma}^{\nu}_{\hspace{2mm}\lambda\rho}v^{\lambda}s^{\mu\rho}
\end{eqnarray*}
are covariant derivatives. Equation (\ref{eq1}) expresses the
coupling of the spin with curvature, and for either a flat
space-time or a spinless particle reduces to the geodesic motion
${\dot{p}}^{\mu}=0$. Also, equation $(\ref{eq2})$ is in fact an
equation showing the balance of the particle and the field angular
momenta. Unlike special relativity, $p^{\mu}$ and $v^{\mu}$ are
not generally proportional to each other. Equations (\ref{eq1})
and (\ref{eq2}) alone are not adequate to give a solution and
three of the components of $s^{\mu\nu}$ remain undetermined by
these equations. This may be regarded as a result of $L$ not being
yet fixed. To remedy this, several supplementary conditions have
been proposed. A widely used condition proposed by Pirani
\cite{pirani} is $v_\mu s^{\mu\nu}=0$. However, it was shown in
the second reference in \cite{t} that this condition leads to
unphysical solutions where the particle can admit helical paths in
flat space-times. Also, it does not uniquely specify  a world line
through the body \cite{wa}. Here, we adopt the following condition
\cite{t}
\begin{equation}
p_{\mu}s^{\mu\nu}=0. \label{eq3}
\end{equation}
This choice does not suffer from the appearance of unphysical
solutions mentioned above and has other advantages discussed in
\cite{di1,di2} and \cite{eh}.  Equations (\ref{eq1}) and
(\ref{eq2}) were derived by Dixon \cite{di1,di2} in a covariant
manner and equation (\ref{eq3}) was adopted as the supplementary
condition. These equations have also been studied from different
angles in \cite{wa}. The post-Newtonian approximation of equations
(\ref{eq1}) and (\ref{eq2}) together with $v_{\mu}s^{\mu\nu}=0$ as
the supplementary equation was studied in \cite{ch}. From Dixon's
equations we  deduce the following relations
\begin{eqnarray*}
p^{\mu}p_{\mu}=\mbox{const.}=-m^2,\hspace{5mm}
\frac{1}{2}s^{\mu\nu}s_{\mu\nu}=\mbox{const.}=s^2.
\end{eqnarray*}
Alternatively, equation (\ref{eq3}) enables us to express the particle's spin
by the four-vector
\begin{eqnarray*}
s^{\mu}{\equiv}\frac{1}{2m}{\eta}^{\mu\nu\lambda\kappa}p_{\kappa}
s_{\nu\lambda}
\end{eqnarray*}
where ${\eta}^{\mu\nu\lambda\kappa}=\frac{1}{\sqrt{-g}}{\epsilon}^
{\mu\nu\lambda\kappa}$ is the alternating tensor with ${\epsilon}^{0123}=1$
and $g$ is the determinant of the metric. It then follows that
$p_{\mu} s^{\mu}=0$ and $s^{\mu}s_{\mu}=s^2$.
Under certain conditions \cite{c}, Dixon's equations may be thought of as the
classical limit of the Dirac equation in curved space-time.

To find the particle's trajectory it is necessary to first find its
four-velocity. Since there is no equations of motion for this purpose, one has
to find the velocity from Dixon's equations in an indirect way. One such
relation is \cite{to}
\begin{equation}
(p\cdot p)v^{\mu}=(p\cdot v)\left( p^{\mu}-\frac{2R_{\sigma\nu\lambda\rho}
p^{\nu}s^{\mu\sigma}
s^{\lambda\rho}}{4p.p-R_{\mu\nu\lambda\kappa}s^{\mu\nu}
s^{\lambda\kappa}}\right). \label{eq4}
\end{equation}
Unlike for $p^{\mu}$ there is no guarantee from the equations of motion
that $v^{\mu}$ remains time-like. In fact Tod {\it et.al.} \cite{to} have
solved Dixon's equations for the case of the motion of a spinning test
particle in the equatorial plane of a Kerr black-hole with the particle's
spin perpendicular
to this plane and found that there are orbits for which $v^{\mu}$ becomes null
or space-like.

As Dixon's equations are reparametrization invariant, we have to
fix a gauge before embarking on finding solutions for them. It is
convenient to choose
\begin{eqnarray*}
\frac{1}{m}p\cdot v=-1.
\end{eqnarray*}
This is the gauge in which the instantaneous zero-momentum and zero-velocity
frames are simultaneous \cite{di2}.
\section{Spinning particles in a weak field}
It is generally difficult to solve equations
(\ref{eq1})-(\ref{eq3}) analytically and to our knowledge, the
only solution that has been found up to now is the previously
mentioned solution by Tod {\it et.al}. However, for the case of
weak fields in a flat background where the metric is of the form
$g_{\mu\nu}=\eta_{\mu\nu}+h_{\mu\nu}$, it is natural to assume
that the change in the particle's velocity and spin are of the
same order as $h_{\mu\nu}$. Therefore, we may decompose the
particle's four-momentum, velocity and spin tensor as
$p^{\mu}=p^{\mu}_0+{\pi}^{\mu}$, $v^{\mu}=v^{\mu}_0+w^{\mu}$ and
$s^{\mu\nu}= s^{\mu\nu}_0+{\sigma}^{\mu\nu}$ with ${\pi}^{\mu}$,
$w^{\mu}$ and ${\sigma}^{\mu}$ being quantities of the same order
as $h_{\mu\nu}$ and $p^{\mu}_0$, $v^{\mu}_0$  and $s^{\mu\nu}_0$
are the initial (before the passage of the wave) four-momentum,
four-velocity and spin tensor of the particle respectively. The
particle's trajectory can then be found from
$\frac{dX^{\mu}(\tau)}{d\tau}=v ^{\mu}$ subject to suitable
initial conditions. Similarly we may express the particle's spin
vector as $s^{\mu}=s^{\mu}_0+{\sigma}^{\mu}$. In terms of these
decompositions, the zeroth order equations become
\begin{equation}
\frac{dp^{\mu}_0}{d\tau}=\frac{ds^{\mu\nu}_0}{d\tau}=0
\label{eqn1}
\end{equation}
\begin{equation}
{p_{\mu}}_0s^{\mu\nu}_0=0 \label{eqn2}
\end{equation}
\begin{equation}
s^{\mu}_0=\frac{1}{2m}{\epsilon}^{\mu}_{\hspace{2mm}\nu\lambda\kappa}v^{\kappa}
_0s^{\nu\lambda}_0. \label{eqn3}
\end{equation}
Equation (\ref{eqn1}) is trivial: in the absence of any field the particle's
four momentum and spin tensor are conserved. Equation (\ref{eqn2}) eliminates
three of the components of $s^{\mu\nu}$ and equation (\ref{eqn3}) may be used
to express $s^{\mu\nu}_0$ in terms of $s^{\mu}_0$. For example, for $p^{\mu}_0=
(m,0,0,0)$ these lead to $s^{0\mu}_0=0$ and $s^{12}_0=s^3_0,s^{13}_0=-s^2_0,s^{
23}_0=s^1_0$.

For the metric $g_{\mu\nu}=g^0_{\mu\nu}+h_{\mu\nu}$, the curvature
tensor and the connection may be written as
$$R_{\mu\nu\kappa\lambda}=
R^{(0)}_{\mu\nu\kappa\lambda}+r_{\mu\nu\kappa\lambda}
\hspace{3mm}\mbox{and}\hspace{3mm}
{\Gamma}^{\mu}_{\hspace{2mm}\nu\kappa}={\Gamma}^{(0
)\mu}_{\hspace{2mm}\nu\kappa}+{\gamma}^{\mu}_{\nu\kappa}.$$
However, the background quantities $R^{(0)}_{\mu\nu
\kappa\lambda}$ and ${\Gamma}^{(0)\mu}_{\hspace{2mm}\nu\kappa}$
vanish for the flat background and the first order form of
equations (\ref{eq1}), (\ref{eq2}) and (\ref{eq4}) reads
\begin{eqnarray}
\frac{d{\pi}^{\mu}(\tau)}{d\tau}&=&-\frac{1}{2}r^{\mu}_{\hspace{2mm}\rho\lambda
\nu}v^{\rho}_0s^{\lambda\nu}_0 -
{\gamma}^{\mu}_{\hspace{2mm}\lambda\nu}v^{\lambda} _0p^{\nu}_0
\label{eq5}
\\
\frac{d{\sigma}^{\mu\nu}(\tau)}{d\tau}&=&-\frac{1}{2m}
r_{\rho\sigma\kappa\lambda}
v^{\nu}_0v^{\sigma}_0s^{\mu\rho}_0s^{\kappa\lambda}_0 -
{\gamma}^{\mu}_{\hspace
{2mm}\lambda\rho}v^{\lambda}_0s^{\rho\nu}_0 -
\mu{\leftrightarrow}\nu \label{eq6}\\
w^{\mu}(\tau)&=&\frac{1}{m}{\pi}^{\mu}(\tau)+\frac{1}{2m^2}r_{\kappa\lambda\rho
\sigma}v^{\lambda}_0s^{\mu\kappa}_0s^{\rho\sigma}_0 \label{eq7}
\end{eqnarray}
with the supplementary equation written as
\begin{eqnarray}
\eta_{\mu\lambda}(s^{\mu\nu}_0{\pi}^{\lambda}(\tau)+p^{\lambda}_0
{\sigma^{\mu\nu}}
(\tau))+h_{\mu\lambda}p^{\lambda}_0s^{\mu\nu}_0=0. \label{eq8}
\end{eqnarray}
The corresponding expression for ${\sigma}^{\mu}$ is
\begin{equation}
{\sigma}^{\mu}(\tau)=\frac{1}{2m}{\epsilon}^{\mu}_{\hspace{2mm}\nu\lambda\rho}
(p^{\rho}_0{\sigma}^{\nu\lambda}+{\pi}^{\rho}s^{\nu\lambda}_0+2p^{\rho}_0s^{
\hspace{1mm}\kappa}_{0\hspace{1mm}\lambda}h^{\nu}_{\hspace{1mm}\kappa}+
p^{\kappa}_0s^{\nu\lambda}_0h^{\rho}_{\hspace{2mm}\kappa}). \label{eq9}
\end{equation}
Therefore, given an initial data set
$v^{\mu}_0=(v^0_0,v^1_0,v^2_0,v^3_0)$ and $s^{\mu}_0
=(s^0_0,s^1_0,s^2_0,s^3_0)$, the orbit of the particle  is given
by
\begin{eqnarray}
X^{\mu}(\tau)=X^{\mu}_0(\tau)+{\chi}^{\mu}(\tau) \label{neq2}
\end{eqnarray}
and its spin components are determined by solving equations
(\ref{eq5})-(\ref{eq8}) subject to suitable initial conditions.
Here, $\frac{dX^{\mu}_0(\tau)}{d\tau}= v^{\mu}_0$ and
$\frac{d{\chi}^{\mu}(\tau)}{d\tau}=w^{\mu}$.
\section{Gravitational waves}
In this section we apply the procedure described in the previous
section to a spinning particle in the field of a weak plane
gravitational wave propagating in the $z$-direction in some
coordinate system. For such a field, the metric may be written as
\cite{g}
\begin{equation}
ds^2=-dt^2+(1+a(t-z))dx^2+(1-a(t-z))dy^2+2b(t-z)dxdy+dz^2
\label{eq99}
\end{equation}
with $a(t-z)=h_{+}\sin(\omega(t-z))$ and
$b(t-z)=h_{\times}\cos(\omega(t-z))$. A wave with linear
polarization is described by $h_{+}=0$ or $h_{\times}=0$ and a
circularly polarized wave is characterized by
$h_{+}={\pm}h_{\times}$.

For this metric, it is straightforward but lengthy to solve the relevant
equations. The resulting expressions for the trajectory of the
particle and the evolution of its spin are too complicated to be included here,
 but are given for $\chi^\mu(\tau)$ in the Appendix. However,
they become manageable for the physically interesting case of the
motion of a particle initially at rest with respect to the
coordinate system in which the metric (\ref{eq99}) is expressed.
In this frame,  $v^0_0=1$,$v^1_0=v^2_0= v^3_0=0$ and an initial
spin orientation may be taken as $(0,s^1_0,s^2_0,s^3_0)$ . For
such a particle the trajectory (with respect to the coordinate
system $(t,x,y,z)$ and the initial conditions
${\chi}^{\mu}(0)=w^{\mu}(0)={\sigma}^{\mu}(0)=0)$ is given by
\begin{eqnarray}
&&\!\!\!\!\!\!\!\!\!t(\tau)=\tau \label{eq10}
\\
&&\!\!\!\!\!\!\!\!\!x(\tau)=-\frac{1}{2m^2}s^2_0(s^3_0{\omega}h_{\times}+
mh_{+})(\omega\tau-\sin
(\omega\tau))+\frac{1}{2m^2}s^1_0(s^3_0{\omega}h_{+}+mh_{\times})
(1-\cos(\omega\tau)) \label{eq11}
\\
&&\!\!\!\!\!\!\!\!\!y(\tau)=-\frac{1}{2m^2}s^1_0(s^3_0{\omega}h_{\times}+
mh_{+})(\omega\tau-\sin
(\omega\tau))-\frac{1}{2m^2}s^2_0(s^3_0{\omega}h_{+}+mh_{\times})
(1-\cos(\omega\tau)) \label{eq12}
\\
&&\!\!\!\!\!\!\!\!\!z(\tau)=\frac{1}{m^2}s^1_0s^2_0{\omega}h_{\times}
(\omega\tau-\sin(\omega\tau)
)+\frac{1}{2m^2}{\omega}h_{+}((s^2_0)^2-(s^1_0)^2)(1-\cos(\omega\tau))
\label{eq13}
\end{eqnarray}
and the equations for spin evolution are
\begin{eqnarray}
&&s^0(\tau)=\frac{\omega}{2m}h_{\times}((s^1_0)^2-(s^2_0)^2)\sin
(\omega\tau)-\frac{\omega}{m}s^1_0s^2_0h_{+}(1-\cos(\omega\tau))\label{eq14}
\\
&&s^1(\tau)=s^1_0-\frac{1}{2}s^1_0h_{+}\sin(\omega\tau)+\frac{1}{2}s^2_0h_
{\times}(1-\cos(\omega\tau)) \label{eq15}
\\
&&s^2(\tau)=s^2_0+\frac{1}{2}s^2_0h_{+}\sin(\omega\tau)+\frac{1}{2}s^1_0h_
{\times}(1-\cos(\omega\tau)) \label{eq16}
\\
&&s^3(\tau)=s^3_0. \label{eq17}
\end{eqnarray}

We can use equations (\ref{eq15})-(\ref{eq17}) to study the
evolution of the particle's spin. From these equations it is clear
that the spin evolution does not depend on $s^3_0$, nor is this
component affected by the wave. Therefore, the initial component
of the spin along the direction of the wave propagation is
unaffected by the wave. The components $s^1(\tau)$ and $s^2(\tau)$
depend on both $s^1_0$ and $s^2_0$ giving rise to the spin vector
precessing about the $z-$axis. The rotation of the spin vector is
very small and as equations (\ref{eq15}-\ref{eq16})  show, they
are of the order of $h_+ s$ or $h_\times s$, where $s$ is the
relevant component of the spin. The appearance of a nonzero
$s^0_0$ component is related to the fact that the particle is not
at rest with respect to this coordinate system.

The particle's trajectory with respect to the frame $(t,x,y,z)$ can be
determined from equations (\ref{eq10})-(\ref{eq13}).
Putting $s^1_0=s^2_0=0$ and $s^3_0=s$ in these equations result in
\begin{eqnarray}
(t(\tau),x(\tau),y(\tau),z(\tau))=(\tau,0,0,0) \label{eqnew1}
\end{eqnarray}
which means that the particle remains at rest at its initial position. This
behaviour is the same as that of a particle without spin. This has
no invariant geometrical meaning and may be interpreted
as follows: the coordinate system $(t,x,y,z)$ is the
one always remaining attached to the particle. The physical effect of the wave
is on the relative motion of the particles. To study this we switch to
$(\overline{t},\overline{x},\overline{y},\overline{z})$, a locally inertial
frame associated with the worldline $x(\tau)=y(\tau)=z(\tau)=0$ given by
\cite{g}
\begin{eqnarray}
\overline{t}&=&t-\frac{1}{4}\frac{\partial{a}}{\partial{t}}(y^2-x^2)-\frac
{\partial{b}}{\partial{t}}xy\label{eq18}\\
\overline{x}&=&x+\frac{1}{2}ax-\frac{1}{2}by-\frac{1}{2}\frac{\partial{a}}
{\partial{t}}xz+\frac{1}{2}\frac{\partial{b}}{\partial{t}}yz\label{eq19}\\
\overline{y}&=&y-\frac{1}{2}ay-\frac{1}{2}bx-\frac{1}{2}\frac{\partial{a}}
{\partial{t}}yz+\frac{1}{2}\frac{\partial{b}}{\partial{t}}xz\label{eq20}\\
\overline{z}&=&z-\frac{1}{4}\frac{\partial{a}}{\partial{t}}(y^2-x^2)-\frac
{\partial{b}}{\partial{t}}xy \label{eq21}
\end{eqnarray}
with $a,b,\partial{a}/\partial{t},$ and $\partial{b}/\partial{t}$ being
evaluated at $x=y=z=0$. In this way, we are able to study the relative
motion of two nearby particles, one with spin and the other without spin.

The motion of a nearby particle without spin initially at the
position $(l_1,l_2,0)$ with respect to this frame was studied in
\cite{p}. The motion of a spinning particle with its spin parallel
to the direction of the wave propagation with respect to this
frame is the same as a particle without spin. Let us now consider
the motion of a particle with its spin perpendicular to the
$z-$axis. We thus set, say, $s^1_0=s$ and $s^2_0=s^3_0=0$ with the
initial position $x=l_1,y=l_2,z=0$. We then obtain from equations
(\ref{eq10})-(\ref{eq13}) and (\ref{eq18})-(\ref{eq21})
\begin{eqnarray}
\overline{t}(\tau)&=&\tau-\frac{{\omega}h_{+}}{4}(l^2_2-l^2_1)
\cos(\omega\tau)+\frac
{{\omega}h_{\times}}{2}l_1l_2\sin(\omega\tau)\label{eq22}\\
\overline{x}(\tau)&=&l_1(1+\frac{h_{+}}{2}\sin(\omega\tau))-\frac{h_
{\times}}{2}l_2\cos(\omega\tau)+\frac{sh_{\times}}{2m}(1-\cos(\omega\tau))
\label{eq23}\\
\overline{y}(\tau)&=&l_2(1-\frac{h_{+}}{2}\sin(\omega\tau))-\frac{h_
{\times}}{2}l_1\cos(\omega\tau)-\frac{sh_{+}}{2m}(\omega\tau-
\sin(\omega\tau))\label{eq24}\\
\overline{z}(\tau)&=&-\frac{{\omega}h_{+}}{4}(l^2_2-l^2_1)
\cos(\omega\tau)-\frac{
{\omega}h_{\times}}{2}l_1l_2\sin(\omega\tau)-\frac{{\omega}h_{+}}{2}
(\frac{s}{m})^2(1-\cos(\omega\tau)). \label{eq25}
\end{eqnarray}
This set of equations determines the trajectory of a spinning
particle initially at rest at the point $(l_1,l_2,0)$ with respect
to a locally inertial frame associated with the origin. In
general, this trajectory is in the form of a  helix whose size is
sensitive to the ratio of the particle's spin to its mass. In
figure 1 the trajectories of a spinning particle and that of a
particle without spin are shown for $\frac{s}{m}\sim 10^{-5}$ m.
We note that in a system of units in which $c=1$, this ratio for
an electron is of the order of $10^{-13}$ m and for an object like
the Earth is $10^{-1}$ m.  For
$\frac{s}{m}{\sim}{\omega}^{-1}{\sim}\lambda$ the spin
contribution to the displacement in the $z-$direction is of the
same order as those for the $x$ and $y-$directions. Also, it is
worth mentioning that for $\frac{s}{m}{\sim}l_1{\sim}l_2$, the
contribution of the terms involving the spin in equations
(\ref{eq23})-(\ref{eq25}) are as important as the terms
representing the motion of a spinless particle. Therefore, the
ratio $\frac{s}{m\lambda}$ seems to play a remarkable role in this
solution.

One may estimate the displacement of a test particle using
equations (\ref{eq23})-(\ref{eq25}). During one period of the
wave, {\it i.e.} $\omega\tau=2\pi$ , a particle initially at
$(l_1,l_2,0)$ would reach the space-time point
\begin{eqnarray*}
\bar{t}&=&\tau-\frac{1}{4}\omega h_+(l_2^2-l_1^2)\\
\bar{x}&=&l_1-\frac{1}{2}h_{\times}l_2\\
\bar{y}&=&l_2-\frac{1}{2}h_{\times}l_1-\pi h_+\frac{s}{m}\\
\bar{z}&=&-\frac{1}{4}h_+\omega(l^2_2-l^2_1).
\end{eqnarray*}
corresponding to a displacement of the order of $hl$ with
$l=\sqrt{l_1^2+l_2^2}$. Choosing the set of parameters $\omega\sim
10^{3}$ Hz, $l_1\sim l_2\sim \frac{s}{m}\sim 10^{-5}$ m and
$h_+\sim h_\times\sim 10^{-21}$, the displacement would be of the
order of $10^{-26}$ m. The above values for $\omega$ and $h_+,
h_\times$ stem from the expectation that for the gravitational
waves accessible to earth based detectors $\omega \sim 1-10^4$ Hz
with a typical amplitude of $h\sim 10^{-21}$ \cite{sc}. This value
for $\omega$ together with $\omega\tau=2\pi$ corresponds to
$\tau\sim 10^{-3}$ sec which is also a typical duration for short
bursts of gravitational waves \cite{b}.

One would hope to be able to detect such gravitational waves by
the future sensitive detectors such as LIGO \cite{sc} currently
under construction. In general, these displacements are small but
could be cumulative and hence likely to be observed.
\begin{figure}
\centerline{ \epsfig{figure=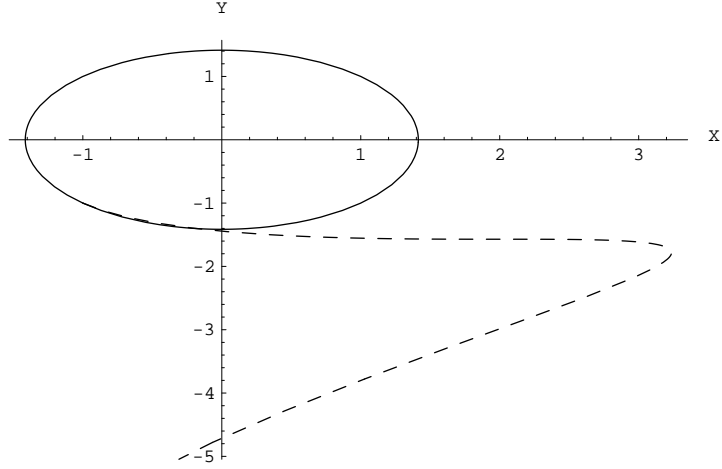, width=10cm} } \caption{The
path of a spinless particle (solid line) and a spinning particle
(dashed line) in the field of a  gravitational wave. They are
originally located at the same position (-1,-1). The values of the
parameters used here are $\omega{\sim}1$ kHz,
$h_+{\sim}h_{\times}{\sim}10^{-21}$, $l_1\sim l_2\sim 10^{-5}$ m
and $\frac{s}{m}{\sim}10^{-5}$ m. For convenience, we have adopted
a system of coordinates given by $X=10^{26}(\overline{x}-10^{-5})$
and $Y=10^{26}(\overline{y}-10^{-5})$. The displacements in the
$Z$ direction are very small and thus are not shown in the
figure.} \label{fig1}
\end{figure}
\section{Conclusions}
We have studied the motion of spinning particles in the field of
gravitational waves. The solutions we have found are free of the
problems mentioned before, namely, the violation of the
normalization condition $v\cdot v=-1$. For a spinless particle,
the velocity and hence its orbit depends on its initial position.
For spinning particles, there is a contribution to the change in
the trajectory from the coupling of the spin to the curvature.
This is independent of the initial position and rather depends on
the initial orientation of the spin. The amount of deviation from
the geodesic path depends on the ratio $\frac{s}{m\lambda}$. For a
certain value of this ratio, the displacement caused by the
spin-gravity coupling could be as important as that caused by the
usual tidal force of gravity waves. Although the particle's
displacement is very small, one may hope to have it amplified in
the course of the wave passage, {\it e.g.}, by a suitable
arrangement of reflectors.

In this paper we have only considered monochromatic waves {\it
i.e.} those with a single $\omega$. In principle it is possible to
generalize this procedure to the case of gravitational waves which
are not monochromatic \cite{v}. Our approximate solutions may also
provide some hints for finding exact solutions to Dixon's
equations for a general plane wave.  For example, we can
investigate whether there are exact solutions to these equations
corresponding to the motion in certain planes with specific spin
orientation. Indeed, equation (\ref{eqnew1}) is such  an exact
solution. This has subsequently been verified in \cite{robin}
where a class of exact solutions along whose path the initial
orientation of the spin of the particle remains unchanged, is
presented. It should also be interesting to study the cumulative
effects of the waves on the motion of particles \cite{p}.

It is worth mentioning that within the context of the perturbative
framework presented here, we have also used  Tulczyjew's
description (a special case of Papapetrou equations together with
Pirani supplementary condition) \cite{morteza}. We have found that
in some special circumstances, {\it e.g.} when the test particle
is initially at rest, the spin evolution is exactly the same for
various supplementary conditions mentioned above. However, in
general  the trajectories turn out to be only slightly different.
The general case of Papapetrou-Pirani will be reported elsewhere
\cite{fut}. \vspace{5mm}\\ {\large{\bf
Acknowledgements}}\vspace{3mm}\\ We would like to thank R. W.
Tucker for introducing this problem to us. M. M. wishes to thank
C. Wang for valuable comments. Both authors acknowledge useful
discussions with S. S. Gousheh.
\section{Appendix}
In this appendix, we give the components of the trajectory of a
test particle \cite{ex} in the field of a gravitational wave
specified by the metric (\ref{eq99}). To make the expressions
manageable, the  following substitutions have been made:
$s_1,s_2,s_3,\rho,\kappa$ stand for
$s^1_0,s^2_0,s^3_0,(\alpha-\delta)
\omega\tau-\sin((\alpha-\delta)\omega\tau)$ and $1-\cos((\alpha
-\delta)\omega\tau)$ respectively with $(v^0_0,v^1_0,v^2_0,v^3_0)=
(\alpha,\beta,\gamma,\delta)$. These components together with
those of the spin (not presented here) satisfy equation
(\ref{eq8}). Also, using equation (\ref{neq2}),  they reduce to
equations (\ref{eq10})-(\ref{eq13}) for the set of initial
parameters discussed in section 4.
\begin{eqnarray*}
\chi^0(\tau)&=&\left\{\rho\left[\omega^2 h_{\times}(s_1^2(\alpha^3
\beta\gamma-3\alpha^2 \beta\gamma\delta-\alpha\beta^3 \gamma
+\alpha\beta\gamma^3+3\alpha\beta\gamma
\delta^2+\beta^3\gamma\delta-
\beta\gamma^3\delta-\beta\gamma\delta^3)\right.\right.\\
&+&s_1s_2(2\alpha^4\delta-\alpha^3\beta^2-\alpha^3\gamma^2-6\alpha^3\delta^2+
\alpha^2\beta^2\delta+\alpha^2\gamma^2\delta+6\alpha^2\delta^3+\alpha\beta^4-
2\alpha\beta^2\gamma^2+ \alpha\beta^2\delta^2\\ &+&
\alpha\gamma^4+\alpha\gamma^2\delta^2-2\alpha\delta^4-\beta^4\delta+2
\beta^2\gamma^2\delta-\beta^2\delta^3-\gamma^4\delta-\gamma^2\delta^3)+
s_1s_3(-\alpha^4\gamma+4\alpha^3\gamma\delta\\
&+&3\alpha^2\beta^2\gamma -
\alpha^2\gamma^3-6\alpha^2\gamma\delta^2-6\alpha\beta^2\gamma\delta+2\alpha
\gamma^3\delta+4\alpha\gamma\delta^3+3\beta^2\gamma\delta^2-\gamma^3\delta^2-
\gamma\delta^4)\\
&+&s_2^2(\alpha^3\beta\gamma-3\alpha^2\beta\gamma\delta+\alpha\beta^3\gamma-
\alpha\beta\gamma^3+3\alpha\beta\gamma\delta^2-\beta^3\gamma\delta+\beta
\gamma^3\delta-\beta\gamma\delta^3)+s_2s_3(-\alpha^4\beta\\
&+&4\alpha^3\beta\delta-\alpha^2\beta^3+3\alpha^2\beta\gamma^2-6\alpha^2
\beta\delta^2+2\alpha\beta^3\delta-6\alpha\beta\gamma^2\delta+4\alpha\beta\delta^3-
\beta^3\delta^2+3\beta\gamma^2\delta^2\\
&-&\beta\delta^4)+s_3^2(-2\alpha^3\beta\gamma+6\alpha^2\beta\gamma\delta-6
\alpha\beta\gamma\delta^2+2\beta\gamma\delta^3))+mh_{+}\omega(s_1(-\alpha^2
\gamma+\alpha\gamma\delta+2\beta^2\gamma)\\ &+&\left. s_2(
-\alpha^2\beta+\alpha\beta\delta+2\beta\gamma^2)+s_3(-2\alpha\beta\gamma+2
\beta\gamma\delta))+2m^2h_{\times}\alpha\beta\gamma\right]+\kappa\left[h_{+}\omega^2(s_1^2(-
\alpha^4\delta\right.\\
&+&\alpha^3\gamma^2+3\alpha^3\delta^2+\alpha^2\beta^2\delta-2\alpha^2\gamma^2
\delta-3\alpha^2\delta^3-2\alpha\beta^2\gamma^2-2\alpha\beta^2\delta^2+\alpha
\gamma^2\delta^2+\alpha\delta^4\\
&+&2\beta^2\gamma^2\delta+\beta^2\delta^3)+s_1s_2(2\alpha\beta^3\gamma-2\alpha
\beta\gamma^3-2\beta^3\gamma\delta+2\beta\gamma^3\delta)+s_1s_3(\alpha^4
\beta-4\alpha^3\beta\delta\\
&-&\alpha^2\beta^3+3\alpha^2\beta\gamma^2+6\alpha^2\beta\delta^2+2\alpha
\beta^3\delta-6\alpha\beta\gamma^2\delta-4\alpha\beta\delta^3-\beta^3
\delta^2+3\beta\gamma^2\delta^2+\beta\delta^4)\\
&+&s_2^2(\alpha^4\delta-\alpha^3\beta^2-3\alpha^3\delta^2+2\alpha^2\beta^2
\delta-\alpha^2\gamma^2\delta+3\alpha^2\delta^3+2\alpha\beta^2\gamma^2-
\alpha\beta^2\delta^2+2\alpha\gamma^2\delta^2
\\
&-&\alpha\delta^4-2\beta^2\gamma^2\delta-\gamma^2\delta^3)+s_2s_3(-\alpha^4
\gamma+4\alpha^3\gamma\delta-3\alpha^2\beta^2\gamma+\alpha^2\gamma^3-
6\alpha^2\gamma\delta^2+6\alpha\beta^2\gamma\delta\\
&-&2\alpha\gamma^3\delta+4\alpha\gamma\delta^3-3\beta^2\gamma\delta^2+
\gamma^3\delta^2-\gamma\delta^4)+s_3^2(\alpha^3\beta^2-\alpha^3\gamma^2-
3\alpha^2\beta^2\delta+3\alpha^2\gamma^2\delta\\
&+&3\alpha\beta^2\delta^2-3\alpha\gamma^2\delta^2-\beta^2\delta^3+\gamma^2
\delta^3))+m{\omega}h_\times(s_1(\alpha^2\beta-\alpha\beta\delta-\beta^3+\beta\gamma^2)+
s_2(-\alpha^2\gamma\\
&+&\left.\left.\alpha\gamma\delta-\beta^2\gamma+\gamma^3)+s_3(\alpha\beta^2-\alpha
\gamma^2-\beta^2\delta+\gamma^2\delta))+m^2h_+(-\beta^2+\gamma^2)\right]\right\}
/(2m^2\omega\alpha(\alpha-\delta))
\\ \\
\chi^1(\tau)&=&\left\{\rho\left[ \omega^2
h_{\times}(s_1^2(\alpha^2 \beta^2\gamma-
2\alpha\beta^2\gamma\delta-\beta^4\gamma+\beta^2\gamma^3+\beta^2\gamma
\delta^2)+s_1 s_2(-\alpha^4\beta
+3\alpha^3\beta\delta\right.\right.\\
&+&\alpha^2\beta^3-\alpha^2\beta\gamma^2-3\alpha^2\beta\delta^2-\alpha\beta^3
\delta-\alpha\beta\gamma^2\delta+\alpha\beta\delta^3-2\beta^3\gamma^2+
2\beta\gamma^4+ 2\beta\gamma^2\delta^2)\\
&+&s_1s_3(2\alpha^2\beta\gamma\delta+2\alpha\beta^3\gamma-2\alpha\beta
\gamma^3-4\alpha\beta\gamma\delta^2-2\beta^3\gamma\delta+2\beta\gamma^3\delta
+2\beta\gamma\delta^3)+s_2^2(\alpha^4\gamma\\
&-&\alpha^3\gamma\delta+\alpha^2\beta^2\gamma-2\alpha^2\gamma^3-\alpha^2
\gamma\delta^2-\alpha\beta^2\gamma\delta+\alpha\gamma^3\delta+
\alpha\gamma\delta^3-\beta^2\gamma^3+\gamma^5+\gamma^3\delta^2)\\
&+&s_2s_3(-\alpha^5+2\alpha^4\delta-\alpha^3\beta^2+3\alpha^3\gamma^2+
2\alpha^2\beta^2\delta-4\alpha^2\gamma^2\delta-2\alpha^2\delta^3+
2\alpha\beta^2\gamma^2-\alpha\beta^2\delta^2
\\
&-&2\alpha\gamma^4-\alpha\gamma^2\delta^2+\alpha\delta^4-2\beta^2\gamma^2
\delta+2\gamma^4\delta+2\gamma^2\delta^3)+s_3^2(-\alpha^4\gamma+2\alpha^3
\gamma\delta-\alpha^2\beta^2\gamma+\alpha^2\gamma^3
\\
&+&2\alpha\beta^2\gamma\delta-2\alpha\gamma^3\delta-2\alpha\gamma\delta^3-
\beta^2\gamma\delta^2+\gamma^3\delta^2+\gamma\delta^4))+mh_{+}\omega(s_1
(\alpha\beta\gamma)+s_2(-\alpha^3+\alpha^2\delta\\
&+&\alpha\gamma^2)+s_3(-\alpha^2\gamma+\alpha\gamma\delta))+\left.m^2h_\times\alpha^2
\gamma\right]+\kappa\left[h_{+}\omega^2(s_1^2(-\alpha^3\beta\delta+\alpha^2\beta
\gamma^2+2\alpha^2\beta\delta^2\right.\\
&+&\alpha\beta^3\delta-\alpha\beta\gamma^2\delta-\alpha\beta\delta^3-2
\beta^3\gamma^2-\beta^3\delta^2)+s_1s_2(-\alpha^4\gamma+\alpha^3\gamma\delta+
3\alpha^2\beta^2\gamma+\alpha^2\gamma^3+\alpha^2\gamma\delta^2
\\&-&\alpha\beta^2\gamma\delta-\alpha\gamma^3\delta-\alpha\gamma\delta^3-
4\beta^2\gamma^3-2\beta^2\gamma\delta^2)+s_1s_3(\alpha^5-2\alpha^4\delta-
\alpha^3\beta^2-\alpha^3\gamma^2\\
&+&2\alpha^2\gamma^2\delta+2\alpha^2\delta^3+4\alpha\beta^2\gamma^2+
3\alpha\beta^2\delta^2-\alpha\gamma^2\delta^2-\alpha\delta^4-4\beta^2
\gamma^2\delta-2\beta^2\delta^3)+s_2^2( -\alpha^4\beta\\
&+&2\alpha^3\beta\delta+3\alpha^2\beta\gamma^2-\alpha^2\beta\delta^2-
2\alpha\beta\gamma^2\delta-2\beta\gamma^4-\beta\gamma^2\delta^2)+s_2s_3
(-4\alpha^3\beta\gamma+6\alpha^2\beta\gamma\delta+4\alpha\beta\gamma^3
\\
&-&4\beta\gamma^3\delta-2\beta\gamma\delta^3)+s_3^2(\alpha^4\beta-2\alpha^3
\beta\delta-2\alpha^2\beta\gamma^2+4\alpha\beta\gamma^2\delta+2\alpha
\beta\delta^3-2\beta\gamma^2\delta^2-\beta\delta^4))\\ &+&
mh_{\times}\omega(s_1(\alpha^3-\alpha^2\delta-\alpha\beta^2)+s_2(-\alpha
\beta\gamma)+s_3(\alpha^2\beta-\alpha\beta\delta))-m^2
h_+\left.\left.\alpha^2\beta\right]\right\}/(2m^2 \omega\alpha^2)
\\ \\
\chi^2(\tau)&=&-\chi^1(\tau\rightarrow-\tau,\beta\leftrightarrow\gamma,s_1\leftrightarrow{s_2})
\\ \\
\chi^3(\tau)&=&\left\{\rho\left[h_{\times}\omega^2(s_1^2(-\alpha^4\beta\gamma+3\alpha^3
\beta\gamma\delta+\alpha^2\beta^3\gamma+\alpha^2\beta\gamma^3-3\alpha^2
\beta\gamma\delta^2-3\alpha\beta^3
\gamma\delta-\alpha\beta\gamma^3\delta+\alpha\beta\gamma\delta^3\right.\right.\\
&+&2\beta^3\gamma\delta^2)+s_1s_2(2\alpha^6-6\alpha^5\delta-3\alpha^4\beta^2-
3\alpha^4\gamma^2+6\alpha^4\delta^2+7\alpha^3\beta^2\delta+7\alpha^3\gamma^2
\delta-2\alpha^3 \delta^3\\
&+&\alpha^2\beta^4+2\alpha^2\beta^2\gamma^2-5\alpha^2\beta^2\delta^2+\alpha^2
\gamma^4-5\alpha^2\gamma^2\delta^2-\alpha\beta^4\delta-6\alpha\beta^2\gamma^2
\delta+\alpha\beta^2\delta^3-\alpha\gamma^4 \delta\\ &+&
\alpha\gamma^2\delta^3+4\beta^2\gamma^2\delta^2)+s_1s_3(\alpha^5\gamma-
4\alpha^4\gamma\delta-\alpha^3\beta^2\gamma-\alpha^3\gamma^3+6\alpha^3\gamma
\delta^2+6\alpha^2\beta^2\gamma\delta\\
&+&2\alpha^2\gamma^3\delta-4\alpha^2\gamma\delta^3-9\alpha\beta^2\gamma
\delta^2-\alpha\gamma^3\delta^2+\alpha\gamma\delta^4+4\beta^2\gamma\delta^3)
+s_2^2(-\alpha^4\beta\gamma+3\alpha^3\beta\gamma\delta\\
&+&\alpha^2\beta^3\gamma+\alpha^2\beta\gamma^3-3\alpha^2\beta\gamma\delta^2
-\alpha\beta^3\gamma\delta-3\alpha\beta\gamma^3\delta+\alpha\beta\gamma
\delta^3+2\beta\gamma^3\delta^2)+s_2s_3(\alpha^5\beta-4\alpha^4
\beta\delta\\
&-&\alpha^3\beta^3-\alpha^3\beta\gamma^2+6\alpha^3\beta\delta^2+
2\alpha^2\beta^3\delta+6\alpha^2\beta\gamma^2\delta-4\alpha^2\beta\delta^3-
\alpha\beta^3\delta^2-9\alpha\beta\gamma^2\delta^2+\alpha\beta\delta^4\\
&+&4\beta\gamma^2\delta^3)+s_3^2(-2\alpha^3\beta\gamma\delta+6\alpha^2\beta
\gamma\delta^2-6\alpha\beta\gamma\delta^3+2\beta\gamma\delta^4))+mh_{+}
\omega(s_1(-\alpha^3\gamma+\alpha^2\gamma\delta+2\alpha\beta^2
\gamma)\\
&+&s_2(-\alpha^3\beta+\alpha^2\beta\delta+2\alpha\beta\gamma^2)+s_3
(-2\alpha^2\beta\gamma+2\alpha\beta\gamma\delta))+\left.2m^2h_\times\alpha^2\beta\right]+
\kappa\left[h_{+}\omega^2(s_1^2( -\alpha^6\right.\\
&+&3\alpha^5\delta+2\alpha^4\beta^2+\alpha^4\gamma^2-3\alpha^4\delta^2-
5\alpha^3\beta^2\delta-2\alpha^3\gamma^2\delta+\alpha^3\delta^3-\alpha^2
\beta^4-\alpha^2\beta^2\gamma^2+ 4\alpha^2\beta^2\delta^2\\
&+&\alpha^2\gamma^2\delta^2+2\alpha\beta^4\delta-\alpha\beta^2\delta^3-
\beta^4\delta^2+\beta^2\gamma^2\delta^2)+s_1s_2(2\alpha\beta^3\gamma\delta-
2\alpha\beta\gamma^3\delta-2\beta^3\gamma\delta^2\\
&+&2\beta\gamma^3\delta^2)+s_1s_3(-\alpha^5\beta+4\alpha^4\beta\delta+
\alpha^3\beta^3+\alpha^3\beta\gamma^2-6\alpha^3\beta\delta^2-4\alpha^2
\beta^3\delta+4\alpha^2\beta\delta^3\\
&+&5\alpha\beta^3\delta^2-3\alpha\beta\gamma^2\delta^2-\alpha\beta\delta^4-
2\beta^3\delta^3+2\beta\gamma^2\delta^3)+s_2^2(\alpha^6-3\alpha^5\delta-
\alpha^4\beta^2-2\alpha^4\gamma^2\\
&+&3\alpha^4\delta^2+2\alpha^3\beta^2\delta+5\alpha^3\gamma^2\delta-
\alpha^3\delta^3+\alpha^2\beta^2\gamma^2-\alpha^2\beta^2\delta^2+
\alpha^2\gamma^4-4\alpha^2\gamma^2\delta^2-2
\alpha\gamma^4\delta\\
&+&\alpha\gamma^2\delta^3-\beta^2\gamma^2\delta^2+\gamma^4\delta^2)+
s_2s_3(\alpha^5\gamma-4\alpha^4\gamma\delta-\alpha^3\beta^2\gamma-
\alpha^3\gamma^3+6\alpha^3\gamma\delta^2+4\alpha^2
\gamma^3\delta\\
&-&4\alpha^2\gamma\delta^3+3\alpha\beta^2\gamma\delta^2-5\alpha\gamma^3
\delta^2+\alpha\gamma\delta^4-2\beta^2\gamma\delta^3+2\gamma^3\delta^3)+
s_3^2(\alpha^3\beta^2\delta-\alpha^3\gamma^2\delta\\
&-&3\alpha^2\beta^2\delta^2+3\alpha^2\gamma^2\delta^2+3\alpha\beta^2\delta^3-
3\alpha\gamma^2\delta^3-\beta^2\delta^4+\gamma^2\delta^4))+m\omega
h_\times (s_1(\alpha^3 \beta-\alpha^2\beta\delta\\
&-&\alpha\beta^3+\alpha\beta\gamma^2)+s_2(-\alpha^3\gamma+\alpha^2\gamma
\delta-\alpha\beta^2\gamma+\alpha\gamma^3)+s_3(\alpha^2\beta^2-\alpha^2
\gamma^2-\alpha\beta^2\delta+\alpha\gamma^2\delta))\\ &+&m^2
h_+\left.\left.
(-\alpha^2\beta^2+\alpha^2\gamma^2)\right]\right\}/(2m^2\omega\alpha^2(\alpha-\delta))
\end{eqnarray*}


\begin{thebibliography}{99}
\bibitem{ex} These calculations have been performed using the
package Excalc in REDUCE.
\end{thebibliography}
\end{document}